\documentstyle[prd,preprint,aps]{revtex}
\tighten
\begin{document}

\reversemarginpar

\title{Building blocks of a black hole}
\author{Jacob D.
Bekenstein\thanks{E-mail:~bekenste@vms.huji.ac.il} and Gilad
Gour\thanks{E-mail:~gour@cc.huji.ac.il}}
\address{Racah Institute of Physics, Hebrew University of
Jerusalem,\\  Givat Ram, Jerusalem~91904, ISRAEL.}

\maketitle

\begin{abstract} What is the nature of the energy spectrum of a
black hole ?  The algebraic approach to black hole quantization
requires the horizon area eigenvalues to be equally spaced.  As
stressed long ago by Mukhanov, such eigenvalues must be
exponentially degenerate with respect to the area quantum
number if one is to understand black hole entropy as reflecting
degeneracy of the observable states.  Here we construct the
black hole stationary states by means of a pair of ``creation
operators'' subject to a particular simple algebra, a slight
generalization of that for a pair of harmonic oscillators.
This algebra reproduces the main features of the algebraic
approach, in particular  the equally spaced area spectrum.  We
then prove rigorously that the $n$-th area eigenvalue is
exactly $2^n$-fold degenerate.  Thus black hole entropy {\it
qua\/} logarithm of the number of states for fixed horizon area
is indeed proportional to that area.
\end{abstract}

\pacs{PACS numbers:~}

\section{Introduction}

Quantum gravity, the interplay of quantum theory with
gravitation  theory, remains one of the most interesting and
challenging topics in theoretical physics today.
Notwithstanding the extant theories\cite{string,loop,canonical}
which purport to represent quantum gravity, there is as yet no
clear and consistent picture of the subject. This is why a
situation involving simultaneously strong gravitational fields
as well as  properties reminiscent of localized particles could
shed light on the construction of the final version of quantum
gravity.  The simplicity of black holes makes them a salient
candidate in this sense. Among the  simplest questions that can
be asked in quantum gravity is what is the nature of the energy
spectrum of a black hole.

One of us noted early that the area of a black hole event
horizon behaves somewhat like a classical adiabatic
invariant~\cite{Bek74} (see also refs.~\cite{BHTrail,Mayo}).
Ehrenfest's principle then suggests that the horizon area
represents a quantum entity with a discrete
spectrum~\cite{Bek74,MG8,Brazil,Erice}.  Further, the fact that
introducing a quantum particle into a Kerr-Newman black hole
carries a minimal ``cost" $\sim\hbar$ of area increase, which
does not depend on the black hole  parameters, suggests that the
spacing between area eigenvalues is
uniform~\cite{bek73,MG8,Brazil,Erice}.  The discrete nature of
the eigenvalue spectrum for the horizon area is also supported
by the loop quantum gravity (see Ashtekar and Krasnov in
Ref.~\cite{loop}), but this last theory suggests a rather
complicated eigenvalue spacing.  If the area spectrum is
equispaced, the classical relation $A=16\pi M^{2}$ ($c=G=1$) for
a Schwarzschild black hole implies the mass spectrum
$M\sim\sqrt{\hbar n}$ for it, where  $n=1,2, \cdots$. This type
of spectrum has subsequently been obtained by many
authors~\cite{APS,mukh86,Authors,berezin}.

The adiabatic invariant approach mentioned is, of course,
heuristic.  Nowdays it is customary to draw conclusions about
observable spectra from an algebra of observables.  The loop
quantum gravity~\cite{loop} indeed seeks to determine the
spectrum of horizon area, among others, from the algebra of
geometric operators in the theory.  A completely different
approach~\cite{MG8,Brazil,Erice,Ann_Arbor} is to assume that
each separate black hole state, which one assumes comes from a
discrete set, is created from a ``black hole vacuum''  $|{\rm
vac}\rangle$ by the operation of a certain unitary operator:
\begin{equation} |njmqs\rangle = \hat{R}_{njmqs}|{\rm
vac}\rangle.
\end{equation}  Here $|njmqs\rangle$ is a one-black hole state
with area $a_{n}$,  angular momentum $j$ ($m$ represents the $z$
component) and charge $q$ (in units of the fundamental charge).
The quantum number $s$  distinguishes between different states
with the same area, charge, angular momentum and its
$z$-component  The algebra of the various $\hat R$ operators
together with the observables, horizon area $\hat A$, charge
$\hat Q$ and angular momentum $\hat {\bf J}$, can be
constructed from symmetry considerations together with the
assumption that any commutator of two $\hat R$'s is linear in
all other $\hat R$'s and  $\hat A$, $\hat Q$ and $\hat {\bf J}$
(which linearity reflects the usual additivity of all these
latter quantities)~\cite{Brazil,Erice,Ann_Arbor}.  Such an
algebra implies  that the spectrum of $\hat{A}$ is equally
spaced  for all charges and angular momenta:
\begin{equation}
 a_{n}=a_{0}n;\qquad n=1,2,3, \cdots,
\label{quant}
\end{equation}  where $a_{0}$ is a positive constant
proportional to $\hbar$.

And where is the  black hole entropy in all this ?  Although the
proportionality of black hole entropy to horizon area can be
inferred solely  by considering the black hole as a macroscopic
system in thermal equilibrium with its surrounding~\cite{GandM},
it is generally agreed today that a crucial test of any proposed
quantum gravity is its ability to recover the above
proportionality from a count of ``internal'' black hole states.
Such derivations of black hole entropy have been proffered in a
number of string related contexts (principally for extreme black
holes)~\cite{count}, by exploiting the asymptotic conformal
symmetry near the horizon\cite{carlip}, and in the loop quantum
gravity~\cite{loop}.  In the algebraic approach on which we
concentrate here, the horizon area eigenvalues are distributed
rather sparsely.  It was first observed by
Mukhanov~\cite{mukh86} (see  also Ref.~\cite{bekmukh}) that the
proportionality of black hole entropy to horizon area is
conditional upon the degeneracy  degree $g_n$, the number of
states
$|njmqs\rangle$ with a common area eigenvalue $a_{n}$, being
given by $g_{n}=k^{n}$, where $k$ is some integer greater than
one.

Heuristic ways of understanding the exponential growth of
degeneracy include the observation that a black hole can
radiatively cascade from the $n$-th level to the ground state
$n=1$ by $2^n$ different paths depending on which area levels it
passes through\cite{DS}, or that it can be raised from the
ground state to level $n$ by steps in $2^n$
ways\cite{bekmukh}.  Another heuristic view is that the
quantization law Eq.~(\ref{quant}) suggests that the horizon
may be regarded as parcelled into $n$ patches of area
$a_0$.  If each can exists in $k$ different quantum states,
then the law $g=k^n$ is immediate\cite{MG8}.
Wheeler\cite{wheeler},  Sorkin\cite{sorkin} and
Kastrup\cite{kastrup} have proposed similar ideas.

The  expectation of an  exponential rise in $g$ is not implicit
in other approaches.  Quantum loop gravity recovers the
connection of the area spectrum with black hole entropy by
predicting a very dense distribution of eigenvalues with little
degeneracy, if any\cite{loop}.  Approaches based on canonical
quantum gravity sometimes predict {\it infinite\/} degeneracy
of sparsely distributed eigenvalues\cite{berezin}.   An
argument within the algebraic approach itself suggests that
$g$ would rise {\it at least as fast\/} as exponentially with
$n$~\cite{MG8,Erice} if it could be assumed that the states
$[\hat{R}_{njmqs}, \hat{R}_{1000s'}]|{\rm vac}\rangle$ with all
allowed $s'$ are independent.  However, at least for the way we
shall construct the $\hat{R}_{njmqs}$ in
Sec.~\ref{sec:algebra}, this last assumption cannot be
maintained.  Formal proof of the law
$g_n=k^n$ has thus been lacking heretofore.

The purpose of the present paper is to show that the
exponential law $g_{n}=k^{n}$ is indeed a consequence of the
algebraic approach if one builds the $R$ operators as products
involving just two kinds of (noncommuting) operators,
$\hat a$'s and $\hat b$'s.  The required algebra of these
``building blocks'' is inferred in Sec.~\ref{sec:algebra} from
very general requirements, and is almost unique.  Then a
systematic procedure is developed in Sec.~\ref{sec:degeneracy}
for counting the number of distinct black hole states created
out of the vacuum by the said operator products.  It yields the
required exponentially rising degeneracy.  The assumptions made
in building the algebra are summarized in
Sec.~\ref{sec:summary}, where possible extensions of the
present ideas are mentioned.

\section{The Algebra}
\label{sec:algebra}
\subsection{Fundamental building blocks}

We start from the intuitive assumption that there exist
one-black hole states.  The normalized vacuum state (no black
hole) is denoted by
$|{\rm vac}\rangle$, and states with nonzero area eigenvalue
$a_{n}$ are denoted by $|n,s\rangle$, where
$s$ is a generic symbol for any additional quantum numbers which
distinguish between all states with common $n$. When the hole
has no angular momentum or charge, we have $s=0,1,2,\cdots
,g_{n}-1$, where $g_{n}$ is the degeneracy of the said
states.  We shall mostly phrase the discussion for this
Schwarzschild black hole case, but our arguments can be
generalized.  As mentioned~\cite{MG8,Brazil,Erice,Ann_Arbor},
operators $\hat{R}_{ns}$ are defined such that  $|n,s\rangle
=\hat{R}_{ns}|{\rm vac}\rangle$. That is,
$\hat{R}_{ns}$ creates a  black hole with area
$a_{n}$ from the vacuum.  One drawback of this scheme is that
there an infinity of creation operators $\hat{R}_{ns}$. It
would be nice to construct them from a small number of more
fundamental ``building blocks'' out of which the whole algebra
of the $\hat R$ operators follows.  At a physical level such
construction, if possible, should illuminate the inner
structure of the black hole.

For simplicity we assume that $g_{1}=2$. Taking
$g_{1}= 3, 4, \cdots $ would change our main result only in some
details.   With $g_{1}=2$ the first area level has two
independent quantum states, say
$|1,0\rangle$ and $|1,1\rangle$.  Let us try identifying the
fundamental building blocks of the algebra  with the $\hat R$
operators for these two states,
\begin{equation}
\hat{a}\equiv\hat{R}_{11}\quad{\rm
and}\quad\hat{b}\equiv\hat{R}_{12},
\label{a=R}
\end{equation}  so that
\begin{equation}   |1,0\rangle\equiv\hat{a}|{\rm
vac}\rangle\quad{\rm and}\quad |1,1\rangle\equiv\hat{b}|{\rm
vac}\rangle.
\label{basis_states}
\end{equation}  By previous
work~\cite{MG8,Brazil,Erice,Ann_Arbor} and Eq.~(\ref{a=R}),
$\hat a$ and $\hat b$ should comply with
\begin{equation}  [\hat{A},\hat{a}]=a_{0}\hat{a}\quad{\rm and}
\quad[\hat{A},\hat{b}]=a_{0}\hat{b},
\label{Aab}
\end{equation}  where $\hat{A}$ is the positive semidefinite
horizon area operator, and $a_0$ is a positive constant with the
dimensions of area.  Eqs.~(\ref{Aab}) are checked by operating
with them on
$|{\rm vac}\rangle$ and taking into account that $\hat A|{\rm
vac}\rangle=0$ because the vacuum contains no horizons.  The
commutators (\ref{Aab}) are taken as axioms here; are they
unique ?

Were one to add to the r.h.s. of the first of Eq.(\ref{Aab}) a
term involving $\hat b$, then $\hat a|{\rm vac}\rangle$ would
no longer be an eigenstate of $\hat A$, thus overturning the
motive in defining $\hat a$.  Similarly for an
$\hat a$ dependent term in the r.h.s. of the second equation.
But it does seem possible, from this point of view, to add to
the r.h.s. in Eq.~(\ref{Aab}) terms of the form $h(\hat A)$,
provided the functions $h$ vanish for zero argument.  However,
by making the redefinitions $\hat a\rightarrow\hat a+h_1(\hat
A)$ and $\hat b\rightarrow\hat b+h_2(\hat A)$, we recover the
original commutators (\ref{Aab}).   Hence for a pair of
building blocks, Eq.~(\ref{Aab}) are the unique choice.  We may
immediately complement them with their conjugates,
\begin{equation}
[\hat{A},\hat{a}^\dagger]=-a_{0}\hat{a}^\dagger\quad{\rm and}
\quad[\hat{A},\hat{b}^\dagger]=-a_{0}\hat{b}^\dagger.
\label{Aab*}
\end{equation}

\subsection{Completing the algebra}
\label{sec:completing}

What are the algebraic relations among the elementary operators
$\hat a$, $\hat b$, $\hat b^\dagger$ and $\hat b^\dagger$ by
themselves ?   One possible approach is that of  Alekseev,
Polychronakos and Smedb\" ack~\cite{APS} who adopt the Cuntz
algebra; in the case of two building blocks this has
$\hat a^\dagger
\hat b=\hat b^\dagger \hat a=0$ and $\hat a^\dagger \hat a=\hat
b^\dagger \hat b=1$.  This algebra yields---trivially---the
exponentially rising degeneracy which is required by the black
hole entropy law.  We wish to stress here, however, the
importance of stating the algebra of the building blocks
exclusively in terms of commutators (as opposed to just
products as in Cuntz's algebra).  This is desirable, not only
because the motivating algebraic
approach~\cite{MG8,Brazil,Erice,Ann_Arbor} is based exclusively
on commutators, but also because of the well known connection
between commutators and Poisson brackets in the
classical-quantum correspondence.  After all, such
correspondence through the adiabatic invariance of horizon area
served as motivation for our approach.

It is easy to verify from the Jacobi identity and
Eq.~(\ref{Aab})-(\ref{Aab*}) that the commutators
$[\hat a^\dagger, \hat b]$, $[\hat b^\dagger,
\hat a]$, $[\hat a^\dagger, \hat a]$ and $[\hat b^\dagger, \hat
b]$ all commute with $\hat A$, but with no other elementary
operator.  Since the elementary operators here are
$\hat a$, $\hat b$, $\hat b^\dagger$, $\hat b^\dagger$ and
$\hat A$, the four mentioned commutators must be functions of
$\hat A$ only:
\begin{eqnarray}
\left[\hat a^\dagger, \hat b\right]&=& \left[\hat b^\dagger,
\hat a\right]^\dagger=f(\hat A)
\label{first}
\\
\left[\hat{a}^\dagger, \hat a\right]&=&\left[\hat{b}^\dagger,
\hat b\right]=F(\hat A),
\label{second}
\end{eqnarray} Here $f={\cal R}(f)+\imath{\cal I}(f)$ may be
complex, but $F$ must be real because $[\hat a^\dagger,\hat a]$
is evidently hermitian.   The assumed equality of the last two
commutators requires  comment. The spirit of the ``no hair
theorems" is that an uncharged and nonrotating black hole has
only one observable degree of freedom, here taken as $\hat A$.
In light of this it seems appropriate to demand
$[\hat{a}^{\dag},\hat{a}]=[\hat{b}^{\dag},\hat{b}]$, since any
asymmetry between the two commutators would provide an extra
observable that distinguished between $\hat a$ and $\hat b$.

Keeping this in mind we see that it should make no difference
physically  if instead of the basis states (\ref{basis_states})
we use linear combinations of them, or equivalently, use linear
combinations $\hat a$ and $\hat b$ to generate basis states.
So suppose we replace the operators
$\hat a$ and $\hat b$ by $\hat a^{'}$ and $\hat b^{'}$
according to
\begin{eqnarray}
\hat a'=\hat a\cos\theta-\hat b\sin\theta
\label{first_theta}
\\
\hat b'=\hat a\sin\theta+\hat b\cos\theta
\label{second_theta}
\end{eqnarray} where $\theta$ is some real angle.  Clearly the
form of the commutators (\ref{Aab})-(\ref{Aab*}) is
unaffected.  However, we now have
\begin{eqnarray}
\left[\hat a'^\dagger, \hat b'\right]&=&{\cal R}(f)\cos
2\theta+\imath{\cal I}(f)
\label{a'b'}
\\
\left[\hat b'^\dagger, \hat a'\right]&=&{\cal R}(f)\cos
2\theta-\imath{\cal I}(f)
\label{b'a'}
\\
\left[\hat a'^\dagger, \hat a'\right]&=&F-{\cal R}(f)\sin
2\theta.
\label{a'a'}
\\
\left[\hat b'^\dagger, \hat b'\right]&=&F+{\cal R}(f)\sin
2\theta
\label{b'b'}
\end{eqnarray} It is apparent that if ${\cal R}(f)\neq 0$, the
form of the algebra varies with $\theta$ contrary to the
principle mentioned.  To eliminate the undesirable dependence on
${\cal R}(f)$, we must demand that ${\cal R}(f)= 0$.

This understood, let us shift the phase of $\hat a$ by
multiplying it by $\imath$.  Clearly this is an allowed
redefinition of $\hat a$ which should have no physical
consequences.  Indeed, it leaves the form of Eqs.~(\ref{Aab}),
(\ref{Aab*}), (\ref{a'a'}) and (\ref{b'b'}) unaffected.
However, it interchanges the roles of
${\cal R}(f)$ and ${\cal I}(f)$ in Eqs.~(\ref{a'b'} ) and
(\ref{b'a'}).  Thus if we afterwards carry out a rotation of the
form (\ref{first_theta})-(\ref{second_theta}), we obvioulsy
find that whenever ${\cal I}(f)\neq 0$, the algebra depends on
$\theta$.  The already explained logic thus forces us to set
${\cal I}(f)=0$ as well.  Overall we must have $f=0$.

Let us operate with Eq.~(\ref{Aab*}) on $|{\rm vac}\rangle$. We
find $\hat A\hat a^{\dag}|{\rm vac}\rangle=-a_0\hat
a^{\dag}|{\rm vac}\rangle$.  But $\hat A$ is a positive
definite operator, so this can only mean that
$\hat a^{\dag}$ anhilates the vacuum.  A similar conclusion
applies to $\hat b^{\dag}$.  It follows from Eqs.~(\ref{first})
and (\ref{second}) that
\begin{eqnarray}
\langle {\rm vac}|\hat{a}^{\dag}\hat{b}|{\rm vac}\rangle&=&
\langle {\rm vac}|\hat{b}^{\dag}\hat{a}|{\rm vac}\rangle=0
\\
\langle {\rm vac}|\hat{a}^{\dag}\hat{a}|{\rm vac}\rangle&=&
\langle {\rm vac}|\hat{b}^{\dag}\hat{b}|{\rm vac}\rangle =F(0).
\end{eqnarray} The first equation tells us that the basis
states $\hat{a}|{\rm vac}\rangle$ and $\hat{b}|{\rm
vac}\rangle$ are orthogonal, which is convenient in what
follows.   To normalize them we require that $F$ be such that
$F(0)=1$.

It will be convenient below to restrict attention to a linear
$F(\hat A)$; but as we explain in Appendix~\ref{appendixA}, our
results about degeneracy remain valid for almost any choice of
function provided $F(x)>0$ for $x>0$.  In light of these remarks
we may focus on the algebra with the form
\begin{eqnarray}
\left[\hat{a}^{\dag},\hat{b}\right]&=&\left[\hat{b}^{\dag},
\hat{a}\right]=0.
\label{adb}
\\
\left[\hat{a}^{\dag},\hat{a}\right]&=&\left[\hat{b}^{\dag},
\hat{b}\right]=1+\alpha\hat{A}
\equiv 1+w\hat{N},
\label{comu}
\end{eqnarray}  where $\alpha$ is an unknown parameter with
dimensions of 1/(area),  $\hat{N}\equiv\hat{A}/a_{0}$ is the
dimensionless area operator and  $w\equiv\alpha a_{0}$. We
shall prove in Sec.~\ref{sec:level2} that necessarily $w>0$. As
a final point here we remark that there is no reason for trying
to express $[\hat a,\hat b]$ in terms of the other basic
operators in the algebra; $[\hat a,\hat b]$ is to be regarded
as a new (certainly nonvanishing) operator.    The relevance of
the algebra (\ref{adb})-(\ref{comu}) for the problem at hand was
first appreciated in conversations of one of us with V.
Mukhanov.  By contrast with the case of Cuntz's algebra, the
establishment of the degeneracy of the area levels for the
algebra (\ref{adb})-(\ref{comu}) is complicated, and requires
the special methods developed in Sec.~\ref{sec:degeneracy}.

\section{Degeneracy of the area levels}
\label{sec:degeneracy}
\subsection{Degeneracy of the $n=2$ area level}
\label{sec:level2}

As mentioned, the first area level, $n=1$ is doubly degenerate.
What is the degeneracy of the $n=2$ states ?  By combining
Eq.~(\ref{Aab}) with the Jacobi identity we find that
\begin{equation} [\hat{A},\hat a\hat a]=2a_0\hat a\hat a;
\quad [\hat{A},\hat b\hat b]=2a_0\hat b\hat b;
\quad [\hat{A},\hat a\hat b]=2a_0\hat a\hat b;
\quad [\hat{A},\hat b\hat a]=2a_0\hat b\hat a.
\label{abcom}
\end{equation}  In view of these, let us define four states
while introducing a new symbol for states:
\begin{equation} |00\rangle\!\rangle\equiv\hat{a}\hat{a}|{\rm
vac}\rangle,\; |01\rangle\!\rangle\equiv\hat{a}\hat{b}|{\rm
vac}\rangle,\; |10\rangle\!\rangle\equiv\hat{b}\hat{a}|{\rm
vac}\rangle\quad{\rm and}\quad
|11\rangle\!\rangle\equiv\hat{b}\hat{b}|{\rm vac}\rangle.
\label{4states}
\end{equation}  In a ket of type $|\ \rangle\!\rangle$ a ``0''
is created by the action of operator $\hat a$ and a ``1'' by
that of $\hat b$.  Operating on $|{\rm vac}\rangle$ with
Eq.~(\ref{abcom}) we find that the above four states are states
with area $2a_0$ corresponding to
$n=2$.  Note that the string of ``0'' and ``1'''s in a state
$|\
\rangle\!\rangle$ is the binary representation of $s$ in our
original notation $|2,s\rangle$ with $s=0,\cdots ,3$.

All states with $n=2$ must be superpositions of the four states
in  Eq.(~\ref{4states}) since there are no other two-operator
products, and it is easy to see, by extending the calculation
entailed in Eq.~(\ref{abcom}), that three-operator product
states, like $\hat a\hat b\hat a|{\rm vac}\rangle$ correspond
rather to $n=3$, and correspondingly larger $n$ for  products of
$n$ operators.  We now prove that the four states are linearly
independent.

Using Eqs.~(\ref{comu}) and (\ref{adb}) one finds that
\begin{eqnarray}
\langle\!\langle 00|00\rangle\!\rangle & = & \langle {\rm
vac}|\hat{a}^{\dag}\hat{a}^{\dag}
\hat{a}\hat{a}|{\rm vac}\rangle= \langle {\rm
vac}|\hat{a}^{\dag}(1+w\hat{N})
\hat{a}|{\rm vac}\rangle+ \langle {\rm
vac}|\hat{a}^{\dag}\hat{a}\hat{a}^{\dag}
\hat{a}|{\rm vac}\rangle=2+w\nonumber\\
\langle\!\langle 10|00\rangle\!\rangle & = &  \langle {\rm
vac}|\hat{a}^{\dag}\hat{b}^{\dag}
\hat{a}\hat{a}|{\rm vac}\rangle= \langle {\rm
vac}|\hat{a}^{\dag}\hat{a}
\hat{a}\hat{b}^{\dag}|{\rm vac}\rangle=0\nonumber\\
\langle\!\langle 10|01\rangle\!\rangle & = &  \langle {\rm
vac}|\hat{a}^{\dag}\hat{b}^{\dag}
\hat{a}\hat{b}|{\rm vac}\rangle= \langle {\rm
vac}|\hat{a}^{\dag}\hat{a}
\hat{b}^{\dag}\hat{b}|{\rm vac}\rangle=1\nonumber\\
\langle\!\langle 01|01\rangle\!\rangle & = &  \langle {\rm
vac}|\hat{b}^{\dag}\hat{a}^{\dag}
\hat{a}\hat{b}|{\rm vac}\rangle= \langle {\rm
vac}|\hat{b}^{\dag}(1+w\hat{N})
\hat{b}|{\rm vac}\rangle=1+w.
\end{eqnarray} By utilizing the symmetry under
$\hat{a}\leftrightarrow\hat{b}$ one can calculate the rest of
the scalar products. Summarizing the scalar products in matrix
form gives
\begin{equation}
\pmatrix{
\langle\!\langle 00|00\rangle\!\rangle & \langle\!\langle
00|01\rangle\!\rangle & \langle\!\langle 00|10\rangle\!\rangle
& \langle\!\langle 00|11\rangle\!\rangle \cr
\langle\!\langle 01|00\rangle\!\rangle & \langle\!\langle
01|01\rangle\!\rangle & \langle\!\langle 01|10\rangle\!\rangle
& \langle\!\langle 01|11\rangle\!\rangle \cr
\langle\!\langle 10|00\rangle\!\rangle & \langle\!\langle
10|01\rangle\!\rangle & \langle\!\langle 10|10\rangle\!\rangle
& \langle\!\langle 10|11\rangle\!\rangle \cr
\langle\!\langle 11|00\rangle\!\rangle & \langle\!\langle
11|01\rangle\!\rangle & \langle\!\langle 11|10\rangle\!\rangle
& \langle\!\langle 11|11\rangle\!\rangle \cr}=\pmatrix{ 2+w & 0
& 0 & 0 \cr 0 & 1+w & 1 & 0 \cr 0 & 1 & 1+w & 0 \cr 0 & 0 & 0 &
2+w
\cr}.
\label{matrix}
\end{equation}

We now show that $w>0$.  Define $|\psi\rangle\equiv
|01\rangle\!\rangle-|01\rangle\!\rangle$.  We have
\begin{equation}
\langle\psi|\psi\rangle=\langle\!\langle
01|01\rangle\!\rangle+\langle\!\langle
10|10\rangle\!\rangle-2\langle\!\langle 01|10\rangle\!\rangle=2w
\end{equation} Of course a minimum requirement is that the norm
of a nontrivial state should be positive.  Hence $w>0$.

The determinant of the matrix in Eq.~(\ref{matrix}) is
$w^4+6w^3+12w^2+8w$. Now were the four states in question
linearly dependent, the above determinant would have to vanish
(a column being a linear combination of the other three).  But
$w>0$, so the four states are linearly independent.  This means
that the degeneracy of the second area level is
$g_{2}=4=2^2$.

This is a good point to indicate why our choice of $F(\hat
A)=1+w\hat A/a_0$ does not restrict the generality of the
conclusions drawn here and below.  With general $F(\hat A)$
satisfying $F(0)=1$ a repetition of the above calculation shows
the nonvanishing entries in the matrix (\ref{matrix}) are
replaced according to $2+w\rightarrow 1+F(a_0)$,
$1+w\rightarrow F(a_0)$ with the unit entry replaced by
$F(0)=1$.   The positivity of the norm
$\langle\psi|\psi\rangle$ would tell us that $F(a_0) >1$.  The
determinant is replaced by $(1+F(a_0))^3 (F(a_0)-1)$ which is
evidently positive.  We again conclude that the four states are
linearly independent.

\subsection{Degeneracy of the $n=3$ and $n=4$ area levels}

For $n=3$ the eight states are
$|3,0\rangle=|000\rangle\!\rangle =\hat{a}\hat{a}\hat{a}|{\rm
vac}\rangle$ and analogously $|3,1\rangle=|001\rangle\!\rangle$,
$|3,2\rangle=|010\rangle\!\rangle$,
$|3,3\rangle=|011\rangle\!\rangle$,
$|3,4\rangle=|100\rangle\!\rangle$,
$|3,5\rangle=|101\rangle\!\rangle$,
$|3,6\rangle=|110\rangle\!\rangle$ and
$|3,7\rangle=|111\rangle\!\rangle$.  Note again that the
sequence of 3-bit ``0'' and ``1'''s is the binary representation
of $s$ in the $|\ \rangle$ form of the ket, while the index
$n=3$ is connected with the fact that the binary representation
is a 3-bit one.  The $8\times 8$ matrix of scalar products of
the eight states $|3,s\rangle$ has been calculated by means of a
dedicated program in {\sl Mathematica\/} which implements the
operator algebra of Eqs.~(\ref{comu}) and (\ref{adb}).  Thus to
calculate $\langle\!\langle 001|010\rangle\!\rangle=\langle{\rm
vac}|\hat b^{\dag}\hat a^{\dag}\hat a^{\dag}\hat a\hat b\hat a
|{\rm vac}\rangle$ one commutes all the $\hat a^{\dag}$ and the
$b^{\dag}$ to the right until they reach $|{\rm vac}\rangle$ and
anhilate it.  The constants produced by the commutations add up
to  $3w+2$.  The full matrix is
\begin{equation}
\left(
\matrix{\scriptstyle 3\,\left( 2 + 3\,w + w^2 \right)
&\scriptstyle 0 &\scriptstyle 0 &\scriptstyle 0 &\scriptstyle 0
&\scriptstyle 0 &\scriptstyle 0 &\scriptstyle 0 \cr  0
&\scriptstyle 2 + 5\,w +
   3\,w^2 &\scriptstyle 2 + 3\,w &\scriptstyle 0 &\scriptstyle 2
+ w &\scriptstyle 0 &\scriptstyle 0 &\scriptstyle 0 \cr 0
&\scriptstyle 2 + 3\,w &\scriptstyle 2 + 3\,w + 2\,w^2
&\scriptstyle 0 &\scriptstyle 2 + 3\,w +
   w^2 &\scriptstyle 0 &\scriptstyle 0 &\scriptstyle 0 \cr 0
&\scriptstyle 0 &\scriptstyle 0 &\scriptstyle 2 + 5\,w + 2\,w^2
&\scriptstyle 0 &\scriptstyle 2 + 3\,w + w^2 &\scriptstyle 2 + w
&\scriptstyle 0 \cr 0 &\scriptstyle 2 +
   w &\scriptstyle 2 + 3\,w + w^2 &\scriptstyle 0 &\scriptstyle
2 + 5\,w + 2\,w^2 &\scriptstyle 0 &\scriptstyle 0 &\scriptstyle
0 \cr 0 &\scriptstyle 0 &\scriptstyle 0 &\scriptstyle 2 + 3\,w
+
   w^2 &\scriptstyle 0 &\scriptstyle 2 + 3\,w + 2\,w^2
&\scriptstyle 2 + 3\,w &\scriptstyle 0 \cr 0 &\scriptstyle 0
&\scriptstyle 0 &\scriptstyle 2 + w &\scriptstyle 0
&\scriptstyle 2 + 3\,w &\scriptstyle 2 + 5\,w +
   3\,w^2 &\scriptstyle 0 \cr 0 &\scriptstyle 0 &\scriptstyle 0
&\scriptstyle 0 &\scriptstyle 0 &\scriptstyle 0 &\scriptstyle 0
&\scriptstyle 3\,\left( 2 + 3\,w + w^2\right) \cr }\right)
\label{matr1}
\end{equation} and its determinant is $46656\,w^6 + 326592\,w^7
+ 1014768\,w^8 + 1842912\,w^9 + 2166588\,w^{10} +
1723356\,w^{11} +
  939681\,w^{12} + 347004\,w^{13} + 83106\,w^{14} +
11664\,w^{15} + 729\,w^{16}$, obviously nonvanishing for
$w>0$.  Hence  the eight $n=3$ states are linearly independent
and the degeneracy of the third area level is thus
$g_{3}=8=2^3$.

For $n=4$ the states are formed by operating a string of four
$\hat a$'s and $\hat b$'s on $|{\rm vac}\rangle$.  Each of these
sixteen states $|4,s\rangle$ with $s=0, 1, \cdots, 15$
corresponds to a state of the form $|\cdots   \rangle\!\rangle$
where the 4-bit binary number equivalent to $s$ reflects the
four operators product used in its construction in accordance
with the equivalence $0\Leftrightarrow\hat a$ and
$1\Leftrightarrow\hat b$.  One can calculate the scalar products
between pairs of states as before; we shall forego the display
of the $16\times 16$ matrix, or its determinant which is also
positive for $w>0$.  Therefore,  the sixteen $n=4$ states are
linearly independent and the degeneracy of the fourth area level
is $g_{4}=16=2^4$.

The pattern is now clear and we proceed to prove analytically
that for a general $n$ area level  the degeneracy is
$g_{n}=2^{n}$.

\subsection{Proof of $2^n$-fold degeneracy of the $n$-th area
level}
\label{sec:proof}

We first define $2^{n}$ states with area eigenvalue $na_{0}$ as
follows:
\begin{equation} |x_{1}x_{2}\cdots
x_{n}\rangle\!\rangle\equiv\hat{x}_{1}\hat{x}_{2}\cdots
\hat{x}_{n} |{\rm vac}\rangle
\label{states}
\end{equation} where $x_{i}=0\;{\rm or}\;1$  and correspondingly
$\hat{x}_{i}$ is either
$\hat{a}$ or $\hat{b}$ ($i=1,2,..,n$). Therefore, there are
exactly $2^{n}$ states.

{\it Theorem}: All the $2^{n}$ states defined in
Eq.~(\ref{states}) are  linearly independent.

This theorem implies that the degeneracy of the $n$th area level
is
$g_{n}=2^{n}$. In order to prove the theorem, we first define an
operator
$\hat{Z}$ which we denote ``quasi-charge",
\begin{equation}
\hat{Z}|x_{1}x_{2}\cdots x_{n}\rangle\!\rangle\equiv
\left(\sum_{i=1}^{n}x_{i}\right) |x_{1}x_{2}\cdots
x_{n}\rangle\!\rangle.
\end{equation} Since $x_{i}=0\;{\rm or}\;1$ the sum
$z\equiv\sum_{i=1}^{n}x_{i}$ counts the  number of times that
$\hat{b}$ appears in the construction of
 $|x_{1}x_{2}\cdots x_{n}\rangle\!\rangle$.

{\it Lemma}: States of like area but different quasi-charge are
orthogonal to each other.

{\it Proof}: We use induction.  For $n=2$ the result is clear
from  Eq.~(\ref{matrix}). Assuming now that it is correct for
$n-1$, we shall prove it for $n$. Let
$|x_{1}x_{2}\cdots x_{n}\rangle\!\rangle$ and
$|x_{1}'x_{2}'\cdots x_{n}'\rangle\!\rangle$ have different
quasi-charges and  consider the following two cases:\\ i)
$x_{1}=x_{1}'$. In this case, by assumption, the state
$|x_{2}\cdots x_{n}\rangle\!\rangle$ is orthogonal to
$|x_{2}'\cdots x_{n}'\rangle\!\rangle$ since they must have
different quasi-charges. Hence,
\begin{eqnarray}
\langle\!\langle x_{1}x_{2}\cdots x_{n}| x_{1}'x_{2}'\cdots
x_{n}'\rangle\!\rangle
 =  \langle\!\langle x_{2}\cdots
x_{n}|\hat{x}_{1}^{\dag}\hat{x}_{1}' | x_{2}'\cdots x_{n}'
\rangle\!\rangle\nonumber\\ = \langle\!\langle x_{2}\cdots
x_{n}|\hat{x}_{1}'\hat{x}_{1}^{\dag} | x_{2}'\cdots
x_{n}'\rangle\!\rangle+[1+(n-1)w]
\langle\!\langle x_{2}\cdots x_{n}| x_{2}'\cdots
x_{n}'\rangle\!\rangle=0,
\end{eqnarray} because the state
$(\hat{x}_{1}'\hat{x}_{1}^{\dag})| x_{2}'\cdots
x_{n}'\rangle\!\rangle$  can evidently be written as a
superposition of states with the same quasi-charge  (and area)
as the state  $| x_{2}'\cdots x_{n}'\rangle\!\rangle$.\\ ii)
$x_{1}\neq x_{1}'$. Without loss of generality, we shall assume
that
$\hat{x}_{1}'=\hat{a}$ and $\hat{x}_{1}=\hat{b}$. If we denote
the  quasi-charge of $|x_{2}\cdots x_{n}\rangle\!\rangle$ by
$z$, then, the quasi-charge of
$| x_{2}'\cdots x_{n}'\rangle\!\rangle$ must be {\it different}
from
$z+1$. Now, when the  operator $\hat{b}^{\dag}\hat{a}$ act on
the state
$| x_{2}'\cdots x_{n}'\rangle\!\rangle$ it preserves the state's
area but decreases its quasi-charge by one. Thus, the state
$\hat{b}^{\dag}\hat{a}| x_{2}'\cdots x_{n}'\rangle\!\rangle$ can
be written as a superposition of states with area $(n-1)a_{0}$
and with a quasi-charge which is different from $z$. By our
assumption $\hat{b}^{\dag}\hat{a}| x_{2}'\cdots
x_{n}'\rangle\!\rangle$ is thus orthogonal to
$|x_{2}\cdots x_{n}\rangle\!\rangle$ and hence
\begin{equation}
\langle\!\langle x_{2}\cdots x_{n}|\hat{b}^{\dag}\hat{a}|
x_{2}'\cdots x_{n}'\rangle\!\rangleð=
\langle\!\langleð x_{1}x_{2}\cdots x_{n}| x_{1}'x_{2}'\cdots
x_{n}'\rangle\!\rangle =0.
\end{equation}  This proves the case $n$; hence by induction
states with different quasi-charge are orthogonal.

The $2^{n}$ states defined in Eq.~(\ref{states}) can be divided
into $n+1$ groups, each characterized by the quasi-charge of
its states: $z=0,1,\cdots ,n$. Thus, the number of states in
the $z$ group  is $ n \choose z$ and the total number of states
with area $na_0$ is $\sum_{z=0}^{n} {n\choose z}=2^{n}$. Since
states with different
$z$ are  orthogonal, it is enough to prove that the
$n\choose z$ states in each $z$ group are all independent. In
the following, states $|x_{1}x_{2}\cdots x_{n}\rangle\!\rangle$
and
$| x_{1}'x_{2}'\cdots x_{n}'\rangle\!\rangle$ with the same $z$
will be denoted by
$|n,z,l\rangle$ and $|n,z,l'\rangle$, respectively, where
$l,l'=1,2,\cdots ,{n\choose z}$.

In order to prove the theorem, it is necessary to know the form
of the scalar product between two general states with the same
quasi-charge $z$. In Appendix~\ref{appendixA} it is shown that
\begin{equation}
\langle n,z,l|n,z,l'\rangle=\sum_{\tilde{p}\in {\cal
P}_{l,l'}}h(\tilde{p}),
\end{equation} where ${\cal P}_{l,l'}$ is the set of
$z!(n-z)!$ permutations (a subset of all the $n!$ permutations
constituting the symmetric, or permutation, group over $n$
objects) that take  string
$x_{1}'x_{2}'\cdots x_{n}'$ representing $|n,z,l\rangle$ into
$x_{1}'x_{2}'\cdots x_{n}'$ representing $|n,z,l'\rangle$. The
function $h(\tilde{p})$ is a specific  one-to-one function that
maps each particular permutation $\tilde{p}$ to a positive
number.

We shall prove by contradiction that the determinant of the
matrix $M^{(n,z)}$ with components $M^{(n,z)}_{ll'}\equiv
\langle n,z,l|n,z,l'\rangle$ is nonvanishing. Let us assume
otherwise.  Then there should be at least one  $n\choose z$
dimensional vector  $\vec{C}\neq 0$ which satisfies
$M^{(n,z)}\vec{C}=0$. This implies that
\begin{equation}
\sum_{l'=1}^{n\choose z}M^{(n,z)}_{ll'}c_{l'}=
\sum_{l'=1}^{n\choose z}\sum_{\tilde{p}\in {\cal P}_{l,l'}}
h(\tilde{p})c_{l'}=0,
\label{zero}
\end{equation} where not all the $c_{l'}$ are zero. Since each
group ${\cal P}_{l,l'}$  contains exactly $z!(n-z)!$
permutations, the sums in  Eq.~(\ref{zero}) contains
$z!(n-z)!\cdot {n\choose z}=n!$ terms. Furthermore,
\begin{equation} {\cal P}_{l,l'}\cap {\cal P}_{l,l''}={\cal
P}_{l',l}\cap {\cal P}_{l'',l}=\{\O\}
\label{empty}
\end{equation} for $l'\neq l''$ because a permutation
$\tilde{p}\in {\cal P}_{l,l'}$ takes the state
$|n,z,l\rangle$ into the state
$|n,z,l'\rangle$, but cannot take
$|n,z,l\rangle$ into $|n,z,l''\rangle$. Note also that
\begin{equation} {\cal P}=\bigcup_{l'=1}^{n\choose z}{\cal
P}_{l',l}=\bigcup_{l'=1}^{n\choose z}{\cal
P}_{l,l'}\qquad\forall
\quad1\leq l\leq {n\choose z},
\label{groups}
\end{equation} where ${\cal P}$ is the symmetric group over $n$
objects. Eq.~(\ref{empty}) and Eq.~(\ref{groups}) will be very
helpful in the following definitions.

Let us define the matrices
$G_{l}$ $(l=1,2,\cdots ,{n\choose z})$,  each of dimension
$z!(n-z)!\times {n\choose z}$. The $k$-th row of
$G_{l}$ is the string of $z!(n-z)!$ randomly ordered distinct
numbers $h(\tilde{p})$ with $\tilde{p}\in {\cal P}_{k,l}$ [as
mentioned, ${\cal P}_{k,l}$ contains $z!(n-z)!$  permutations].
Note that Eqs.~(\ref{empty})- (\ref{groups}) imply that each
matrix
$G_{l}$ contains exactly all the $n!$ terms
$h(\tilde{p})$ with $\tilde{p}\in {\cal P}$.

We now construct the matrix $H_{1}\equiv G_{1}\cup G_{2}\cup
\cdots \cup G_{n\choose z}$, of dimension  $n!\times {n\choose
z}$ by taking its columns to be the columns of all the
$G_{l}$'s in the given order. By enlarging the $n\choose z$
dimensional  vector $\vec{C}$ into the $n!$ dimensional vector
\begin{equation}
\vec{C}_{\rm enlarged}=(\underbrace{c_{1},c_{1},\cdots
,c_{1}}_{z!(n-z)!},
\underbrace{c_{2},c_{2},\cdots ,c_{2}}_{z!(n-z)!},\cdots ,
\underbrace{c_{n\choose z},c_{n\choose z},\cdots ,c_{n\choose
z}}_{z!(n-z)!}),
\end{equation} one can show that Eq.~(\ref{zero}) is equivalent
to
$H_{1}\vec{C}_{\rm enlarged}=0$.

The matrix $G_{l}^{(m)}$, where $m=1,2,\cdots ,z!(n-z)!$, is
obtained after performing $m$ cyclic permutations to the columns
of
$G_{l}$.  Thus for $m=1$ the second column is replaced by the
first, the third by the second, etc. For $m=2$ the first is
replaced by the third, etc.  Hence, all the
$z!(n-z)!$ matrices $H_{m}\equiv G_{1}^{(m-1)}\cup
G_{2}^{(m-1)}\cup\cdots \cup G_{n\choose z}^{(m-1)}$ satisfy
$H_{m}\vec{C}_{\rm enlarged}=0$. Finally, the square matrix $H$
of dimension $n!\times n!$ is defined such that its  first
$n\choose z$ rows are given by $H_{1}$, the second $n\choose z$
rows by $H_{2}$ and so on. Therefore, it is clear that also
$H\vec{C}_{\rm enlarged}=0$.

In each row and each column of $H$ $all$ the $n!$ numbers
$h(\tilde{p})$ appear.  Hence, by writing out the set
$\{h(\tilde{p})|\tilde{p}\in {\cal P}\}$ as
$h_{1},h_{2},\cdots ,h_{n!}$,  we find that
\begin{equation} H=\pmatrix{ h_{1} & h_{2} & h_{3} & \cdot &
\cdot & \cdot & h_{n!-1} & h_{n!} \cr  h_{n!} & h_{1} & h_{2} &
\cdot & \cdot & \cdot & h_{n!-2} & h_{n!-1} \cr   h_{n!-1} &
h_{n!} & h_{1} & \cdot & \cdot & \cdot & h_{n!-3} & h_{n!-2}
\cr
\cdot & \cdot & \cdot & \cdot & \cdot & \cdot & \cdot & \cdot
\cr
\cdot & \cdot & \cdot & \cdot & \cdot & \cdot & \cdot & \cdot
\cr
\cdot & \cdot & \cdot & \cdot & \cdot & \cdot & \cdot & \cdot
\cr h_{2} & h_{3} & h_{4} & \cdot & \cdot & \cdot & h_{n!} &
h_{1}
\cr}
\label{matr2}
\end{equation} where we have rearranged the rows in $H$
(changing the orders of the rows  in $H$ does not affect the
equation $H\vec{C}_{\rm enlarged}=0$).

Let us now recall the $n!$-th roots of unity:
\begin{equation}
\varepsilon_{m}=\exp\left(i\frac{2\pi}{n!}m\right);\qquad m=1,
2, \cdots , n!
\label{epsilon}
\end{equation}  It may be checked that the eigenvectors of $H$
are
\begin{equation}
\vec{e}_{k}\equiv
(\varepsilon_{1}{}^{k},\varepsilon_{2}{}^k,\cdots
,\varepsilon_{m}{}^k ,\cdots ,\varepsilon_{n!}{}^k); \qquad k=1,
2, \cdots , n!
\end{equation}  with corresponding eigenvalues
\begin{equation}
\lambda_{k}=h_{1}+h_{2}\varepsilon_{1}{}^k+h_{3}\varepsilon_{2}{}^k
+\cdots +h_{n!}\varepsilon_{n!-1}{}^k.
\label{lam}
\end{equation}  Because $\varepsilon_{1}{}^{n!}=1$, and the
$h_m$ are positive,
$\lambda_{n!}>0$. It can be shown (Appendix~\ref{appendixB})
that $\lambda_k\neq 0$ also for $k<n!$.  Thus the determinant of
$H$ is $not$ zero. This contradicts our tentative assumption
that there exists a vector $\vec{C}\neq 0$ such that
$M^{(n,z)}\vec{C}=0$ because this is equivalent to assuming
that $H\vec{C}_{\rm enlarged}=0$.   Thus the matrix $M^{(n,z)}$
must have nonvanishing determinant, which proves that all the
$2^{n}$ states defined in Eq.~(\ref{states}) are linearly
independent, as claimed.   In the last paragraph of
Appendix~\ref{appendixA} we explain why this result remains
unchanged if our choice $F(\hat{A})=1+w\hat{A}/a_{0}$ is
replaced by almost any other function which is positive for
positive argument.

\section{Summary and Conclusions}
\label{sec:summary}

The equally spaced area spectrum of a stationary black hole
raises the question of the degeneracy of area states.  An old
argument by Mukhanov\cite{mukh86} suggests that the degeneracy
should rise exponentially with the area quantum number $n$ if
the black hole entropy is to be understood as the logarithm of
the number of ``microstates'' per state with definite
observable parameters, e.g. area, charge, etc.  What algebra of
operators would be conducive to such behavior ?  We have here
assumed that the generic black hole state is created by
operating on the vacuum with a string of ``raising'' operators
of just two kinds, $\hat a$ and $\hat b$ (building blocks).
Assuming the commutator of either  $\hat a$ or
$\hat b$ with the area operator $\hat A$ is proportional to
itself, this construction explains the equispaced area
spectrum.  By further assuming that the algebra of the  $\hat
a$ and $\hat b$ operators and their adjoints is exclusively
defined by commutators, and that no physical consequences
follow when  $\hat a$ and $\hat b$ are redefined as linear
combinations of themselves, we have singled out the algebra,
and we have shown that it leads to the degeneracy law $g_n=2^n$
which is of the type needed to explain the black hole entropy
as a reflection of area eigenstate degeneracy.

The above described arguments do not depend on the exact nature
of the stationary black hole: spherical or rotating, neutral or
electrically charged.  It would evidently be interesting to
associate with the $\hat{a}$ and $\hat{b}$ operators some
angular momentum and/or charge, and so build up specifically
Reissner-Nordstr\" om and Kerr black hole states. We know that
some degeneracy accrues to systems with angular momentum by
virtue of rotational symmetry: states with definite area
$na_0$, squared angular momentum $j(j+1)\hbar^2$ and charge $q$
should comprise substates differing only by the $z$-component
of angular momentum.  Thus the black hole degeneracy might be
expected to depend not only on the area quantum number
$n$, but also on the angular momentum $j$.  However, according
to the first law of thermodynamics, the black hole entropy is a
function of the horizon area alone\cite{GandM} and, therefore,
so should the degeneracy. This  implies that the spectrum of the
horizon area of a black hole must depend on all of $n,\;j$ and
$q$. This argument is consistent with the  result from
canonical quantum gravity obtained by Barvinsky, Das and
Kunstatter~\cite{BDK01} for the area  spectrum of charged black
holes and gives a further motivation for our  algebraic
approach.

\section{Acknowledgments} JDB thanks T. Damour for an invitation
to the Institute des Hautes Etudes Scientifiques where the
problem solved here was first examined with V. Mukhanov, who is
to be thanked for his inspired suggestions.  We both thank
Alexios Polychronakos for useful remarks.  This research was
supported by grant No. 129/00-1 of the Israel Science
Foundation to JDB and by a Clore Foundation doctoral fellowship
to GG.

\appendix

\section{scalar product of two generic states}
\label{appendixA}
Let  $|x_{1}x_{2}\cdots x_{n}\rangle\!\rangle$ and $|
x_{1}'x_{2}'\cdots x_{n}'\rangle\!\rangle$ be two states with
the same area (later we shall assume the same $z$ also).  Their
scalar product is
\begin{equation}
\langle\!\langle x_{1}x_{2}\;\cdots x_{n}|x_{1}'x_{2}'\;\cdots
\;x_{n}'\rangle\!\rangle=
\langle{\rm vac}|\hat{x}_{n}^{\dag}\hat{x}_{n-1}^{\dag}\cdots
\hat{x}_{1}^{\dag}
\hat{x}_{1}'\hat{x}_{2}'\cdots \hat{x}_{n}'|{\rm vac}\rangle,
\end{equation} which we rewrite by succesively moving $\hat
x_1^{\dag}$ all the way to the right using
Eqs.~(\ref{comu})-(\ref{adb}):
\begin{eqnarray}   &=&\delta_{x_{1},x_{1}'}[1+(n-1)w]\langle{\rm
vac}|\hat{x}_{n}^{\dag}\cdots
\hat{x}_{2}^{\dag}\hat{x}_{2}'\cdots \hat{x}_{n}'|{\rm
vac}\rangle+
\langle{\rm vac}|\hat{x}_{n}^{\dag}\cdots
\hat{x}_{2}^{\dag}\hat{x}_{1}'
\hat{x}_{1}^{\dag}\hat{x}_{2}'\cdots \hat{x}_{n}'|{\rm
vac}\rangle\nonumber\\ & = &
\delta_{x_{1},x_{1}'}[1+(n-1)w]\langle\!\langle x_{2}\cdots
x_{n}|x_{2}'\cdots x_{n}'
\rangle\!\rangle+\delta_{x_{1},x_{2}'}[1+(n-2)w]\langle\!\langle
x_{2}\cdots x_{n}|x_{1}'x_{3}' \cdots
x_{n}'\rangle\!\rangle\nonumber\\ &+&\cdots
+\delta_{x_{1},x_{n}'}\langle\!\langle x_{2}\cdots
x_{n}|x_{1}'\cdots x_{n-1}'
\rangle\!\rangle.
\end{eqnarray}  We have used the fact that $x_1^{\dag}|{\rm
vac}\rangle=0$.  Now we move $\hat x_2^{\dag}$ all the way to
the right
\begin{eqnarray}  & &\langle\!\langle x_{1}x_{2}\cdots
x_{n}|x_{1}'x_{2}'\cdots x_{n}'\rangle\!\rangle=
\delta_{x_{1},x_{1}'}[1+(n-1)w] \Bigl\{ \delta_{x_{2},x_{2}'}
[1+(n-2)w]
\langle\!\langle x_{3}\cdots x_{n}|x_{3}'\cdots
x_{n}'\rangle\!\rangle\nonumber\\ & + &
\delta_{x_{2},x_{3}'} [1+(n-3)w]
\langle\!\langle x_{3}\cdots x_{n}|x_{2}'x_{4}'x_{5}'\cdots
x_{n}'\rangle\!\rangle +\cdots +\delta_{x_{2},x_{n}'}
\langle\!\langle x_{3}\cdots x_{n}|x_{2}'\cdots
x_{n-1}'\rangle\!\rangle
\Bigr\} \nonumber\\  & + &\delta_{x_{1},x_{2}'}[1+(n-2)w]
\Bigl\{
\delta_{x_{2},x_{1}'}  [1+(n-2)w]\langle\!\langle x_{3}\cdots
x_{n}|x_{3}'\cdots x_{n}'\rangle\!\rangle\nonumber\\ & +
&\delta_{x_{2},x_{3}'} [1+(n-3)w]
\langle\!\langle x_{3}\cdots x_{n}|x_{1}'x_{4}'x_{5}'\cdots
x_{n}'\rangle\!\rangle  +\cdots +\delta_{x_{2},x_{n}'}
\langle\!\langle x_{3}\cdots x_{n}|x_{1}'x_{3}'x_{4}'\cdots
x_{n-1}'\rangle\!\rangle \Bigr\}
\nonumber\\  &+&\cdots +
\delta_{x_{1},x_{n}'}\Bigl\{ \delta_{x_{2},x_{1}'} [1+(n-2)w]
\langle\!\langle x_{3}\cdots x_{n}|x_{2}'\cdots
x_{n-1}'\rangle\!\rangle\nonumber\\ &+&\delta_{x_{2},x_{2}'}
[1+(n-3)w]
\langle\!\langle x_{3}\cdots x_{n}|x_{1}'x_{3}'x_{4}'\cdots
x_{n-1}'\rangle\!\rangle +\cdots +\delta_{x_{2},x_{n-1}'}
\langle\!\langle x_{3}\cdots x_{n}|x_{1}'x_{2}'\cdots
x_{n-2}'\rangle\!\rangle \Bigr\}.
\label{scalar}
\end{eqnarray}

Thus generically  the scalar product is a sum of many  (actually
$n!$) terms. One example is
$$ [1+(n-1)w]
[1+(n-2)w]\cdots[1+(n-n)w]\delta_{x_{1},x_{1}'}\delta_{x_{2},x_{2}'}
\cdots\delta_{x_{n},x_{n}'}.
$$  obtained by converting by the aforesaid means the first
term within the  first curly brackets in Eq.~(\ref{scalar}).
Other examples include the term
$$ [1+(n-1)w]
[1+(n-3)w][1+(n-3)w][1+(n-4)w]\cdots[1+(n-n)w]\delta_{x_{1},x_{1}'}
\delta_{x_{2},x_{3}'}\delta_{x_{3},x_{2}'}
\cdots\delta_{x_{n},x_{n}'}
$$  resulting from expansion of the second term within the same
brackets, and the term
$$ [1+(n-1)w]
[1+(n-2)w]\cdots[1+(n-n)w]\delta_{x_{1},x_{1}'}
\delta_{x_{2},x_{2}'}\cdots\delta_{x_{n},x_{n}'}.
$$  coming from the last term within the last curly brackets of
Eq.~(\ref{scalar}).  Summing up, the  scalar product has the
following form:
\begin{eqnarray}
\langle\!\langle x_{1}x_{2}&\cdots &x_{n}|x_{1}'x_{2}'\;\cdots
\;x_{n}'\rangle\!\rangle\nonumber\\ & = & \sum_{\tilde{p}\in
{\cal P}}[1+(n-i_{1})w][1+(n-i_{2})w]\cdots
[1+(n-i_{n})w]\delta_{x_{1},x_{p_{1}}'}\delta_{x_{2},x_{p_{2}}'}
\cdots\delta_{x_{n},x_{p_{n}}'}\;,
\label{product}
\end{eqnarray} where ${\cal P}$ is the (symmetric) group of all
$n!$ permutations  $\tilde{p}\equiv (p_{1},p_{2},\cdots
,p_{n})$ of the objects labelled by $1, 2, \cdots, n$ and
$i_{1}, i_{2},\cdots ,i_{n}$ are
$n$ integers satisfying $1\leq i_{1}\leq n,\;2\leq i_{2}
\leq n,\; \cdots \;,n-1\leq i_{n-1}\leq n,\;i_{n}=n$. Hence,
there are exactly $n!$ sets of $i_{1}, i_{2},\cdots ,i_{n}$ and
each  permutation $\tilde{p}$ can be regarded as associated
with a single set  $i_{1}(\tilde{p}), i_{2}(\tilde{p}),\cdots
,i_{n}(\tilde{p})$.

Eq.~(\ref{product}) supplies an alternative proof of the lemma
of section III: the scalar product of  two states with
different quasi-charge must be zero. This is because
$\delta_{x_{1},x_{p_{1}}'}\delta_{x_{2},x_{p_{2}}'}
\cdots\delta_{x_{n},x_{p_{n}}'}=0$
for all $\tilde{p}$. Therefore, we shall restrict ourselves to
states with a fixed  quasi-charge $z$.

As mentioned in Sec.~\ref{sec:proof}, there are $n\choose z$
states with the same $z$, we shall  denote them by
$|n,z,l\rangle\!\rangle\equiv|x_{1}x_{2}\cdots
x_{n}\rangle\!\rangle$ (or
$|n,z,l'\rangle\!\rangle\equiv|x_{1}'x_{2}'\cdots
x_{n}'\rangle\!\rangle$), where
$l,l'=1,2,\cdots ,{n\choose z}$. Furthermore, the product
$\delta_{x_{1},x_{p_{1}}'}\delta_{x_{2},x_{p_{2}}'}\cdots
\delta_{x_{n},x_{p_{n}}'}$ is not zero for exactly $z!(n-z)!$
permutations. Thus, we shall denote by ${\cal P}_{l,l'}$ the
group of $z!(n-z)!$ permutations that contribute to the scalar
product of  $|n,z,l\rangle\!\rangle$ with
$|n,z,l'\rangle\!\rangle$. Using these notation, we can write
the scalar product of two states in a compact form:
\begin{equation}
\langle\!\langle n,z,l|n,z,l'\rangle\!\rangle=\sum_{\tilde{p}\in
{\cal P}_{l,l'}}
\prod_{k=1}^{n}\left[1+(n-i_{k}(\tilde{p}))w\right]\equiv
\sum_{\tilde{p}\in {\cal P}_{l,l'}}h(\tilde{p}).
\end{equation}  Notice that all $h(\tilde{p})$ are positive and
different.

In the paper, the above explicit expression for
$h(\tilde{p})$ is not  used. Therefore, the choice of
$F(\hat{A})=1+w\hat{A}/a_{0}$ does not restrict the
generality of the conclusions drawn. For general $F(\hat{A})$,
$h(\tilde{p})$ can be written as
\begin{equation}
h(\tilde{p})=\prod_{k=1}^{n}F[a_{0}(n-i_{k}(\tilde{p}))].
\end{equation}
The only restriction on the function $F(x)$, apart from
the required $F(0)=1$ (see Sec.~\ref{sec:completing}),
is that all the $n!$ numbers $\{h(\tilde{p})\}$ appearing in
the above equation be positive and distinct.  Almost any
function with $F(x)>0$ for $x>0$ will do.

\section{$\lambda_k\neq 0$}
\label{appendixB}

We shall prove here that $\lambda_{k}$ defined in
Eq.~(\ref{lam}) is nonzero for $k\neq n!$ (in
Sec.~\ref{sec:proof} we have remarked that $\lambda_{n!}>0$).
The proof is by contradiction.  Let us assume one or more of
the $\lambda_k$ with $k<n!$ vanish, so that $\det H=0$.  Thus,
if we interchange rows or columns of $H$, the determinant
remains zero.  Let us reorder the columns so that the upper row
is composed of positive numbers in order of increasing
magnitude, which we shall again denote by $h_{1}, h_{2},
\cdots\;h_{n!}$.  By appropriately exchanging rows we can bring
the new matrix,
$H'$, to look exactly like that in Eq.~(\ref{matr2}) with
$0<h_{1}<h_{2}<\cdots <h_{n!}$. Obviously $\det H'=0$. We shall
denote the eigenvalues of $H'$ by $\lambda_k'$.

According to Eq.~(\ref{epsilon}),
$\varepsilon_{m}{}^k\varepsilon_{1}{}^k=\varepsilon_{m+1}{}^k$.
Thus
\begin{equation}
\lambda_{k}'(1-\varepsilon_{1}{}^k)=h_{1}+(h_{2}-h_{1})\varepsilon_{1}{}^k
+(h_{3}-h_{2})\varepsilon_{2}{}^k+\cdots
+(h_{n!}-h_{n!-1})\varepsilon_{n!-1}{}^k -h_{n!}.
\label{llam}
\end{equation} Taking the absolute value of Eq.~(\ref{llam}) we
find that
\begin{equation}  |\lambda_{k}'(1-\varepsilon_{1}{}^k)|\geq
h_{n!}-\left|h_{1}+(h_{2}-h_{1})
\varepsilon_{1}{}^k+(h_{3}-h_{2})\varepsilon_{2}{}^k+\cdots
+(h_{n!}-h_{n!-1})
\varepsilon_{n!-1}{}^k\right|,
\label{ggg}
\end{equation} where we have used the fact that $|x-y|\geq
\left||x|-|y|\right|$ for any two complex numbers
$x$ and $y$.  In writing Eq.~(\ref{ggg}) we have taken into
account that its r.h.s. cannot be negative since in light of
the inequality  $|x+y|\leq |x|+|y|$,
\begin{eqnarray} &\;& \left|h_{1}+(h_{2}-h_{1})
\varepsilon_{1}{}^k+(h_{3}-h_{2})\varepsilon_{2}{}^k+\cdots
+(h_{n!}-h_{n!-1})
\varepsilon_{n!-1}{}^k\right|\nonumber\\ & \leq &
h_{1}+(h_{2}-h_{1}) +(h_{3}-h_{2})+\cdots +
(h_{n!}-h_{n!-1})=h_{n!}.
\end{eqnarray}

We now show that the r.h.s. of  Eq.~(\ref{ggg}) cannot vanish.
For if it vanished, the definitions
$\alpha_{1}\equiv h_{1}/h_{n!}$ and
$\alpha_{m}\equiv (h_{m}-h_{m-1})/h_{n!}$ for $2\leq m\leq n!$
would imply that
\begin{equation}
\left|\sum_{m=1}^{n!}\alpha_{m}\varepsilon_{m}{}^k\right|=1
\label{robin}
\end{equation} so that
$\sum_{m=1}^{n!}\alpha_{m}\varepsilon_{m}{}^k=\exp(i\gamma)$
with $\gamma$ real.  Therefore, we would have
\begin{equation}
\sum_{m=1}^{n!}\alpha_{m}\exp\left[i\left(\frac{2\pi k}{n!}m
-\gamma\right)\right]=1.
\label{condition}
\end{equation}  On the other hand, by definition all
$\alpha_{m}$ are positive and
\begin{equation}
\sum_{m=1}^{n!}\alpha_{m}=1.
\label{gilrob}
\end{equation}  Thus Eq.~(\ref{condition}) can hold only if
$k=0$ or $k=n!$ and
$\gamma=0$ (mod $2\pi$).  We conclude that for $k\neq 0$ and
$k\neq n!$, the r.h.s. of Eq.~(\ref{ggg}) is necessarily
positive; the equation then shows that $\lambda_{k}'\neq 0$ for
$k=1, 2, \cdots, n!-1$. From Eq.~(\ref{lam}) it again follows
that $\lambda_{n!}'>0$ because the $n!$-th power of all
$\varepsilon_m$ is unity.  Thus, contrary to assumption,
$\det H=\det H'$ cannot vanish.  The contradiction tells us
that all $\lambda_m$ of the original matrix $H$ are
nonvanishing.

\end{document}